\pdfoutput=1
\documentclass[aps,prx,reprint,floatfix]{revtex4-1}

\usepackage[utf8]{inputenc}
\usepackage{amsmath}

\usepackage{graphicx}
\usepackage[usenames, dvipsnames]{color}
\usepackage{float}
\usepackage{array}
\usepackage{color}
\graphicspath{{Figures/}}
\usepackage{xspace}
\usepackage{amssymb}

\newcommand{\ket}[1]{| #1 \rangle}

\newcommand{\kHz}{\mathrm{kHz}}
\newcommand{\MHz}{\mathrm{MHz}}
\newcommand{\GHz}{\mathrm{GHz}}
\newcommand{\cm}{\mathrm{cm}}
\newcommand{\nm}{\mathrm{nm}}
\newcommand{\mK}{\mathrm{mK}}

\newcommand{\us}{\mathrm{\mu}\mathrm{s}}
\newcommand{\s}{\mathrm{s}}
\newcommand{\ms}{\mathrm{ms}}
\newcommand{\ns}{\mathrm{ns}}
\newcommand{\kb}{k_\mathrm{B}}

\newcommand{\n}{n_g}

\newcommand{\Ej}{E_J}
\newcommand{\Ec}{E_C}
\newcommand{\ejoec}{E_J/E_C}

\newcommand{\xqp}{x_\mathrm{QP}}

\newcommand{\nlb}{\nolinebreak}

\newcommand{\geven}{\ket{0,e}}
\newcommand{\godd}{\ket{0,o}}
\newcommand{\eeven}{\ket{1,e}}
\newcommand{\eodd}{\ket{1,o}}
\newcommand{\venv}{V_\mathrm{env}}
\newcommand{\pcorr}{\langle P(0)P(\tau)\rangle}

\begin{document}

\widetext

\title{Direct Dispersive Monitoring of Charge Parity in Offset-Charge-Sensitive Transmons}

\author{K.~Serniak}
\email{kyle.serniak@yale.edu}
\affiliation{Department of Applied Physics, Yale University, New Haven, CT 06520, USA}
\author{S.~Diamond}
\affiliation{Department of Applied Physics, Yale University, New Haven, CT 06520, USA}
\author{M.~Hays}
\affiliation{Department of Applied Physics, Yale University, New Haven, CT 06520, USA}
\author{V.~Fatemi}
\affiliation{Department of Applied Physics, Yale University, New Haven, CT 06520, USA}
\author{S.~Shankar}
\affiliation{Department of Applied Physics, Yale University, New Haven, CT 06520, USA}
\author{L.~Frunzio}
\affiliation{Department of Applied Physics, Yale University, New Haven, CT 06520, USA}
\author{R.~J.~Schoelkopf}
\affiliation{Department of Applied Physics, Yale University, New Haven, CT 06520, USA}
\author{M.~H.~Devoret}
\email{michel.devoret@yale.edu}
\affiliation{Department of Applied Physics, Yale University, New Haven, CT 06520, USA}

\begin{abstract}
A striking characteristic of superconducting circuits is that their eigenspectra and intermode coupling strengths are well predicted by simple Hamiltonians representing combinations of quantum circuit elements.
Of particular interest is the Cooper-pair-box Hamiltonian used to describe the eigenspectra of transmon qubits, which can depend strongly on the offset-charge difference across the Josephson element. 
Notably, this offset-charge dependence can also be observed in the dispersive coupling between an ancillary readout mode and a transmon fabricated in the offset-charge-sensitive (OCS) regime. 
We utilize this effect to achieve direct, high-fidelity dispersive readout of the joint plasmon and charge-parity state of an OCS transmon, which enables efficient detection of charge fluctuations and nonequilibrium-quasiparticle dynamics.
Specifically, we show that additional high-frequency filtering can extend the charge-parity lifetime of our device by two orders of magnitude, resulting in a significantly improved energy relaxation time~$T_1\sim200~\mu\mathrm{s}$.
\end{abstract}

\maketitle
\section{Introduction}
The basic building blocks of quantum circuits---e.g.~capacitors, inductors, and nonlinear elements such as Josephson junctions~\cite{Josephson1962} and electromechanical transducers~\cite{Teufel2011}---can be combined and arranged to realize device Hamiltonians engineered for specific tasks~\cite{Vool2017}.
This includes a wide variety of superconducting qubits for quantum computation~\cite{Bouchiat1998,Mooij1999,Chiorescu2003,Koch2007,Manucharyan2009}, quantum-limited microwave amplifiers~\cite{Castellanos-Beltran2009,Bergeal2010,Macklin2015}, and frequency converters for quantum signal routing~\cite{Andrews2014}.
These circuits can be probed using standard rf measurement techniques and understood within the theoretical framework of circuit quantum electrodynamics (cQED)~\cite{Blais2004}, which has been used to accurately predict energy levels and intermode coupling strengths in novel and complex circuits~\cite{Janvier2015,Smith2016,Kou2017}.
Arguably the most well-studied quantum circuit is the capacitively shunted Josephson junction~\cite{Bouchiat1998,Koch2007}, which is parameterized by the ratio of the Josephson coupling energy $\Ej$ to the charging energy $\Ec$. This circuit is typically operated in either the Cooper-pair box~($\ejoec\approx1$)~\cite{Bouchiat1998} or transmon~($\ejoec\gtrsim50$)~\cite{Koch2007} extremes of offset-charge sensitivity.
We will focus on circuits that fall in the range between these two extremes.
There, the characteristic plasmonic eigenstates (which we will refer to as plasmon states) of the circuit can be superpositions of many charge states, like a usual transmon, but with measurable offset-charge dispersion of the transition frequencies between eigenstates, like a Cooper-pair box.
This defines what we refer to as the offset-charge-sensitive (OCS) transmon regime.\\
\indent Devices fabricated in the OCS regime are particularly useful for investigations of interesting mesoscopic phenomena.
For example, these devices can be used to probe deviations from the typical sinusoidal Josephson current-phase relation, which will change the offset-charge dependence of circuit eigenenergies and transition matrix elements~\cite{Ginossar2014,Yavilberg2015}. 
Additionally, this offset-charge dependence in devices with standard Al/AlOx/Al junctions can facilitate sensitive measurements of environmental charge noise and quasiparticle dynamics~\cite{Schreier2008,Riste2013,Serniak2018}.
This is important because the performance of superconducting devices, especially qubits, can be limited by dissipation due to nonequilibrium quasiparticles (QPs)~\cite{Lutchyn2005,Martinis2009,Catelani2011}.
The fact that the observed ratio of these nonequilibrium QPs to Cooper pairs~($\xqp \approx\nlb 10^{-8}$~to~$10^{-5}$~\cite{Aumentado2004,Segall2004,Naaman2006,Shaw2008,Martinis2009,Vool2014,Nsanzineza2014,Wang2014,DeVisser2014,Gustavsson2016,Taupin2016,Serniak2018}) is many orders of magnitude greater than would be expected in low-temperature experiments~($\sim20~\mK$) remains an unsolved mystery.
Nonetheless, given this observed phenomenological range of $\xqp$, the natural combination of cQED and BCS theory~\cite{Bardeen1957} leads to quantitative modeling of QP-induced dissipation that has shown good agreement with experiments~\cite{Lutchyn2005,Martinis2009,Catelani2011,Lenander2011,Wang2014}.
Recent work has demonstrated that  the effects of QPs can even be distinguished from other sources of dissipation in OCS transmons~\cite{Riste2013,Serniak2018}. 
These experiments were able to correlate qubit transitions with changes in the charge-parity of the circuit: a signature of QPs interacting with the qubit~\cite{Catelani2014}. 
This development has provided a foundation for experiments aiming to mitigate QP-induced dissipation and identify the generation mechanisms of nonequilibrium QPs~\cite{Bespalov2016,Catelani2019}.\\
\indent In this article, we present a new, efficient method to monitor the charge parity of an OCS transmon.
This method takes advantage of significant hybridization between the higher-excited plasmon states in an OCS transmon and an ancillary readout mode, resulting in a charge-parity-dependent shift of that readout-mode frequency, even when the transmon is in its ground plasmon state.
We leverage this effect to perform direct, high-fidelity dispersive readout of the joint plasmon and charge-parity state of an OCS transmon over a wide range of offset-charge configurations.
This is in contrast to previous experiments that monitored the charge parity of OCS transmons by relying on state transitions induced by coherent pulses~\cite{Riste2013,Serniak2018}.
The measured charge-parity-dependent dispersive shifts agree with the predictions of quantum circuit theory~\cite{Manucharyan2012,Zhu2013,Smith2016}, and we show that this readout scheme provides a straightforward probe of QP tunneling rates across the OCS transmon Josephson junction. 
This idea of a parametric susceptibility can be extended to study other sources of decoherence, such as charge and flux noise, in cQED systems.\\
\indent Finally, we apply this technique to quantify the effect of high-frequency filtering on quasiparticle dynamics in transmons. 
Measuring the exact same device as in Ref.~\cite{Serniak2018}, we find that improved filtering of radiation with frequency of order $2\Delta/h$ (where $\Delta$ is the superconducting energy gap) increases the timescale between QP-tunneling events by almost two orders of magnitude to $\approx6~\ms$. 
We observe an equilibrium excited-state population $\mathcal{P}_1^\mathrm{eq} \approx 1.5\%$ and an average energy-relaxation time~$T_1 \approx 210~\us$, which agrees with the predictions of Ref.~\cite{Serniak2018}. In this regime of reduced $\xqp$, QPs are \emph{not} a dominant dissipation mechanism in our OCS transmon device.
\section{Hamiltonian of an OCS Transmon}
Transmons are constructed by shunting a Josephson junction with a large capacitance to achieve a charging energy $\Ec$ that is much smaller than the Josephson coupling energy $\Ej$, such that the transition frequency between the ground and first-excited state~($\omega_{01}$) is \emph{greater} than that between the first- and second-excited states~($\omega_{12}$).
Fig.~1(a) shows a general circuit schematic for this type of device coupled to an environment with fluctuating charges represented by a noisy voltage source $V_\mathrm{env}$ that imposes a dimensionless offset charge $\n=\nlb C_gV_\mathrm{env}/2e$ across the junction (note the factor of~$2e$, which references the charge of a Cooper pair). 
Though~$\n$ drifts stochastically, there are often long periods ($\sim$~minutes) of offset-charge stability.
To describe quasiparticle dynamics in this circuit, we employ the familiar Cooper-pair-box Hamiltonian with a slight generalization:
\begin{equation}
    \hat{H}_\mathrm{CPB}=4E_C\left(\hat{n}-\n+\frac{P-1}{4}\right)^2-E_J\cos\hat{\varphi}.
\end{equation}
The Hamiltonian $\hat{H}_\mathrm{CPB}$ contains two operators: $\hat\varphi$ is the superconducting phase difference across the junction and $\hat{n}$ is the number of Cooper pairs that have traversed the junction. The discrete parameter $P=\pm1$ is the charge parity of the circuit (the parity of the total number of \emph{electrons} that have traversed the junction). 
We choose the convention that if no electrons have tunneled, there is zero contribution of the parity term to the Hamiltonian, and thus~``even"~(``odd") corresponds to~$P=+1$~($P=-1$).
Tunneling of a single QP will switch~$P$, which affects the energy spectrum as if~$\n$~was shifted by~$1/2$, creating two manifolds of energy eigenstates indexed by $P$~[Fig.~1(b)].
Coherent Cooper-pair tunneling does not change charge parity.
\begin{figure}
	\centering
    \includegraphics[width=\columnwidth]{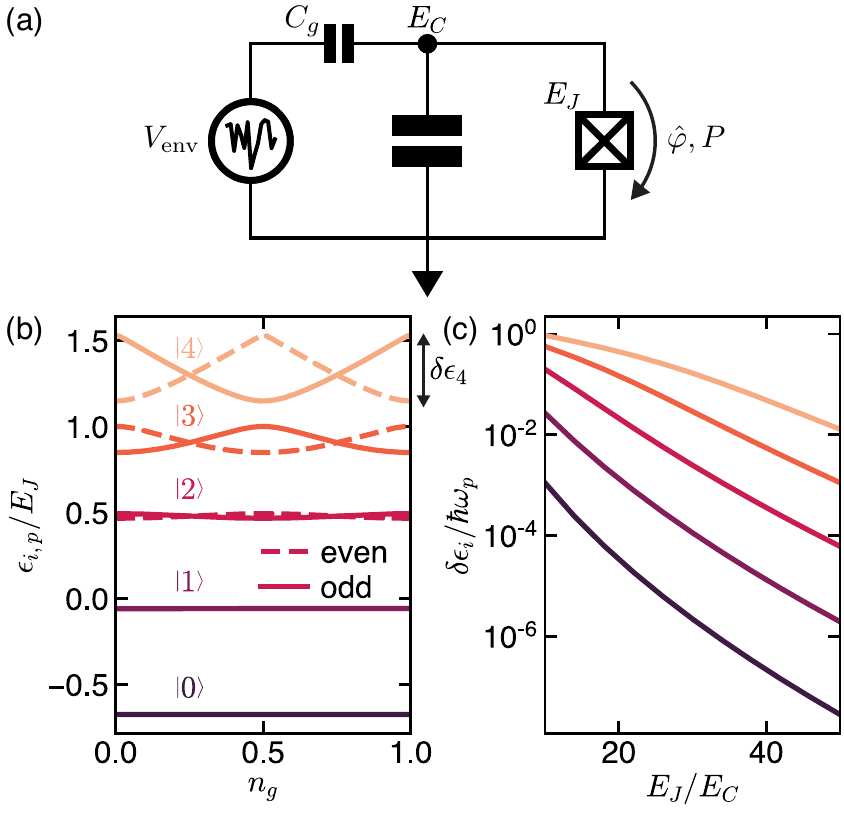} 
	\caption{Offset-charge dispersion of OCS-transmon eigenstates.
	(a)~Circuit diagram of a Cooper-pair box/transmon coupled to charges in the environment.  
    Fluctuating charges in the environment produce noisy reduced charge offset~\mbox{$\n = C_g \venv/2e$,} where $C_g$ is an effective gate capacitance and $e$ is the electron charge. The symbols $\Ej$ and $\Ec$ refer to the Josephson and charging energies, respectively, while $\hat\varphi$ and~$P$ denote the difference in superconducting phase across the junction and the number parity of QPs that have tunnelled across the junction, respectively. 
	(b)~Eigenenergies $\epsilon_{i,p}$ of the Cooper-pair-box Hamiltonian with $E_J/E_C=17$, as a function of $\n$, normalized by $E_J$. 
	Solid (dashed) lines indicate the manifold of states corresponding to odd (even) charge parity.
	(c)~Maximum charge dispersion $\delta \epsilon_i$ of the five lowest energy levels, normalized by the plasma frequency $\omega_p=\sqrt{8E_JE_C}/\hbar$.
	}
\end{figure}\\
\indent The eigenstates of an OCS transmon are indexed by two discrete labels:~$i$ denotes the plasmon-excitation number and~$p$ denotes the charge parity.
For readability, we will indicate~$i$ numerically~(0,1,2...) and~$p$ with label ``$e$" or ``$o$," for ``even" and ``odd" charge parity, respectively.
The eigenenergies~$\epsilon_{i,p}(\n)$ corresponding to our device with $\ejoec=17$ are shown in Fig.~1(b).
In the transmon limit, the presence of two charge-parity manifolds is typically neglected because the maximum charge dispersion of the energy levels $\delta\epsilon_i=|\epsilon_{i,e}(0)-\epsilon_{i,o}(0)|$ decreases exponentially with $\sqrt{\ejoec}$~\cite{Koch2007}~[Fig.~1(c)] and the splitting of the lowest energy levels (those relevant for coherent manipulation in quantum computing architectures) is overcome by other sources of dephasing at the $\sim10~\kHz$ level~\cite{Gambetta2006,Wang2019}. 
\section{Charge-Parity-Dependent Dispersive Shifts}
The strength of the OCS transmon-readout mode coupling will vary with $\n$.
An OCS transmon coupled to a single linear readout mode is described by the Hamiltonian~\cite{Blais2004,Koch2007}
\begin{equation}
    \hat{H}=\hat{H}_\mathrm{CPB}+\hbar\omega_r \hat{a}^\dagger\hat{a}+\hbar{g}\hat{n}\left(\hat{a}+\hat{a}^\dagger\right).
\end{equation}
Here, $\omega_r$ is the bare readout mode frequency, $g$ is the capacitive coupling rate between the OCS transmon and the readout mode, and $\hat{a}$ is the bosonic annihilation operator for excitations in the bare readout mode. 
In the dispersive regime, the coupling term $\hbar g\hat{n}(\hat{a}+\hat{a}^\dagger)$ produces a transmon-state-dependent shift $\chi_{i,p}$ of the readout mode frequency relative to $\omega_r$.
Such dispersive shifts are the basis for qubit readout in cQED.
Up to second order in perturbation theory, $\chi_{i,p}$ can be written~\cite{Manucharyan2012}
\begin{equation}
    \chi_{i,p} = g^2\sum_{j\neq i}\frac{2\omega_{ij,p}\left|\left\langle j,p|\hat{n}|i,p\right\rangle\right|^2}{\omega_{ij,p}^2-\omega_\mathrm{r}^2},
\end{equation}
\begin{figure}
	\centering
    \includegraphics[width=\columnwidth]{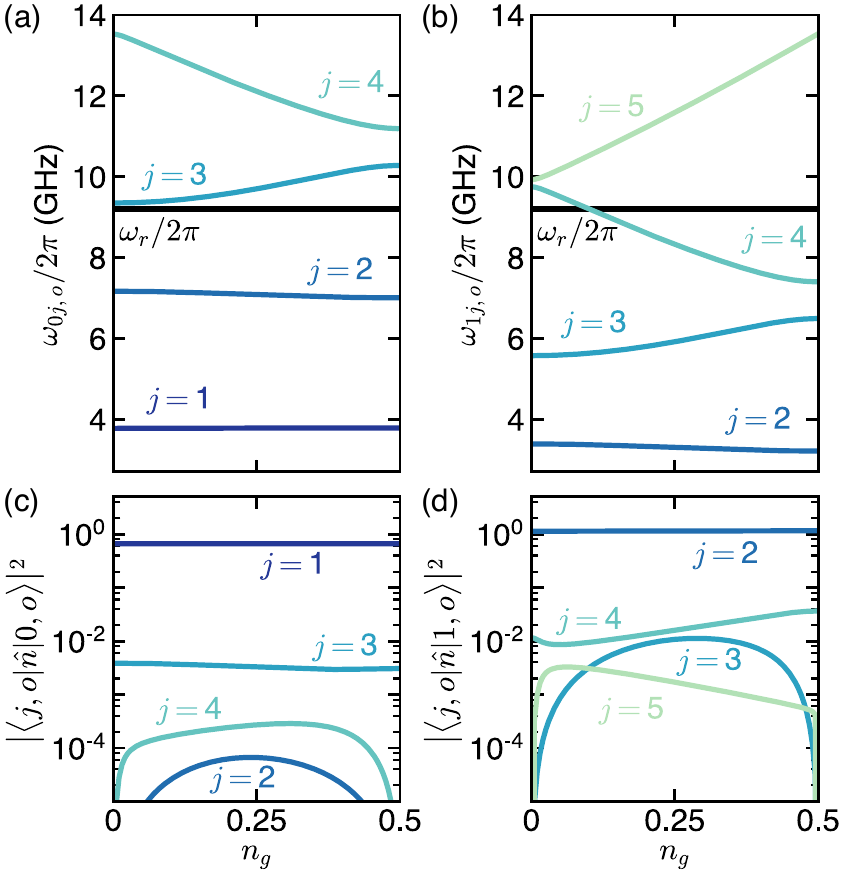} 
	\caption{Theoretically calculated OCS transmon-resonator spectrum as a function of $\n$.
    Plasmon transition frequencies out of the ground state (a) and first-excited state (b) of an OCS transmon with $\ejoec = 17$ with odd charge parity.
	In this parameter regime, the detuning between~$\omega_{03,o}$ and the resonator frequency~$\omega_r$ varies by a factor of~$\approx8$ as a function of the parameter~$\n$. Additionally,~$\omega_{14,o}$ crosses~$\omega_r$ near~$\n=0.1$. 
	Matrix elements of the transmon charge operator for transitions out of the ground~(c) and first excited~(d) states with ``odd" charge parity. 
	These matrix elements are finite and relevant for calculating the transmon-resonator dispersive shifts in our devices. 
	In the transmon limit of large~$\ejoec$, matrix elements between non-nearest-neighbor states will be suppressed.}
\end{figure}\\
which is valid for $g\left|\left\langle j,p|\hat{n}|i,p\right\rangle\right|\ll(\omega_{ij,p}-\omega_r)$.
Here, $\omega_{ij,p}$ is the transition frequency between transmon states $\ket{i,p}$ and $\ket{j,p}$. For a harmonic oscillator, only the charge matrix elements $\langle j,p|\hat{n}|i,p\rangle$ coupling nearest-neighbor $i$ and $j$ are nonzero.
In a traditional weakly anharmonic transmon, $\chi_{i,p}$ is well approximated by including only nearest-neighbor terms, except in the rare case where a transmon transition is nearly resonant with the readout mode.
In the more anharmonic OCS regime, charge dispersion of the transmon levels~[Fig.~1(c)] can significantly change the detuning of transition frequencies from the readout mode~[Fig.~2(a,~b)].
In addition, the charge matrix elements coupling non-nearest neighbor transmon states become important~[Fig.~2(c,~d)].
We calculate these quantities by numerical diagonalization in the charge~($\hat{n}$) basis. 
It is worth noting that the dominant matrix elements are relatively insensitive to $\n$.
It is only necessary to consider transitions out of the two lowest-energy transmon eigenstates because the steady-state thermal population of higher levels can be neglected in the regime where~$\hbar\omega_{01}\gg \kb T$.
For visual clarity, we plot only the transitions belonging to the ``odd" charge-parity manifold; the ``even" transition frequencies and matrix elements are mirror symmetric about the degeneracy point~$\n=0.25$.\\
\indent The parameters chosen for Figs.~1~and~2 reflect the experimental device that will be discussed in the next sections: $\Ej/h = 6.14~\GHz$, $\Ec/h = 356~\MHz$, and~$\omega_r/2\pi\approx9.202~\GHz$. Notice that, in this parameter regime,~$\omega_{03,o}(\n)$ comes close to the bare readout frequency at~$\n=0$, and that~$\omega_{14,o}(\n)$ crosses the resonator mode frequency near~$\n=0.1$. These lead to substantial changes of the dispersive shifts of the readout resonator as a function of $\n$. Given a readout mode frequency in the typical range of cQED systems, only modest tuning of $\Ej$ and $\Ec$ is required to observe the dispersive effects discussed above, as long as the ratio~$\ejoec$ is sufficiently low.
\section{Experimental setup}
The experiments presented here were performed on the exact same device as in Ref.~\cite{Serniak2018}.
To recapitulate, an OCS transmon is coupled to a Al 3D waveguide cavity~\cite{Paik2011} and the transmon state is read out through a standard rf input/output chain by detecting the amplitude and phase of a signal reflected from the input of the cavity.
During the six months since the experiments reported in Ref.~\cite{Serniak2018}, the device was stored in air at room temperature.
In this time, the Al-AlOx-Al Josephson junction ``aged''~\cite{Pop2012}, decreasing $E_J$ such that $\ejoec =\nlb 23\rightarrow\nlb17$ and $\overline{\omega_{01}}/2\pi=\nlb4.4004~\GHz\rightarrow\nlb3.7837~\GHz$.
Here,~$\overline{\omega_{01}}=\nlb|\omega_{01,e}(\n)+\nlb \omega_{01,o}(\n)|/2$ for any value of~$\n$, and is also the time-average of both~$\omega_{01,e}$ and~$\omega_{01,o}$ assuming ergodic fluctuations of~$\n$.
This shift produced a corresponding change of the maximum charge dispersion of the 0-1 transition $\delta \omega_{01}(0)/2\pi=\nlb1.6~\MHz\rightarrow\nlb6.3~\MHz$.
Crucially for our experiment, the charge dispersion of higher-excited states~($i\geq\nlb2$) is greatly increased such that there is significant variation of the OCS transmon-resonator mode dispersive shift with~$\n$.
The rf lines and filters~[Appendix~A] are similar to those shown in the Supplemental Material of Ref.~\cite{Serniak2018}.
There are a few differences, namely an additional Al shield surrounding the sample and improved rf low-pass filtering on the input/output line inside of this shield.
We attribute an observed reduction of $\xqp$ to the latter, which will be discussed in the next section.
\begin{figure}
	\centering
    \includegraphics[width=\columnwidth]{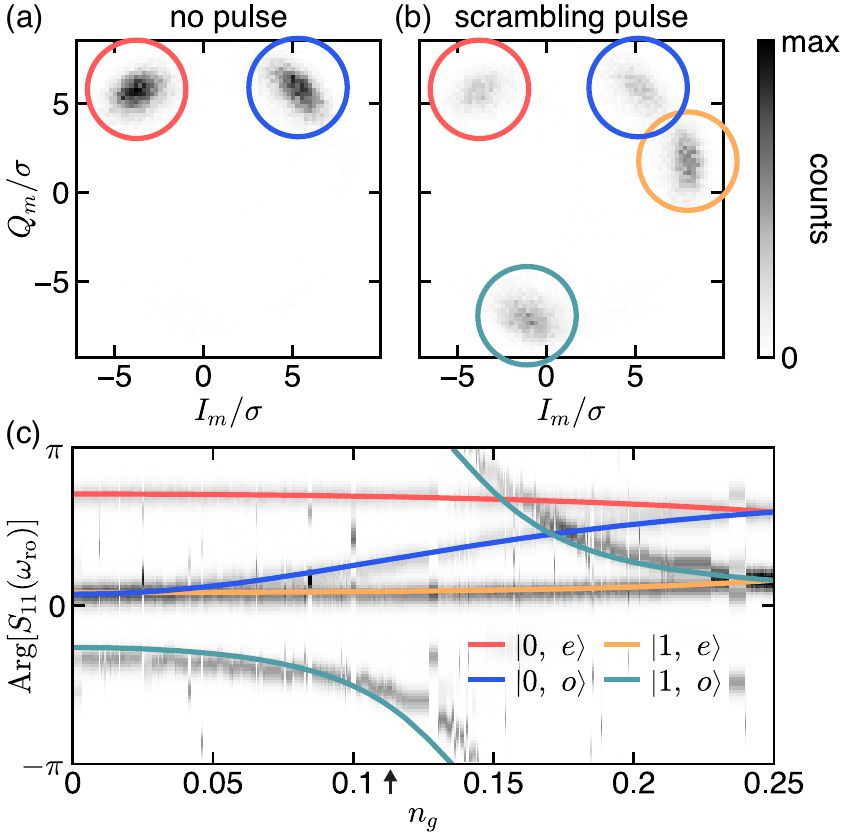} 
	\caption{Direct dispersive readout of the joint plasmon and charge-parity state of an OCS transmon. (a) Histogram in the complex plane of $2\times10^4$ sequential shots separated by~$200~\us$ and integrated for~$4.16~\us$, normalized by~$\sigma$, the standard deviation of the measurement distributions obtained by projecting onto the~$I_m$ axis and fitting to a sum of two Gaussian functions~[Fig.~4(b)]. Measurements circled in red (blue) are assigned to denote the state~$\geven$~($\godd$). 
	(b) Histogram obtained under the same conditions as in (a), but with a pulse applied before each measurement to scramble the qubit state. 
	Measurements circled in yellow (green) are assigned to denote the state $\eeven$ ($\eodd$).
	(c) Histograms of the phase of the readout signal sorted by~$\n$.
	Due to symmetry of the eigenspectrum, our measurement maps all values of $\n$ into the range $[0,1/4]$.
	The histograms in~(a) and~(b) correspond to the data marked by the black arrow at~$\n=0.11$.
    }
\end{figure}
\section{Results}
\indent Figs.~3-5 describe the main experimental result of this article: the direct-dispersive measurement of the joint plasmon and charge-parity state of an OCS transmon. 
Due to the charge dispersion of the OCS transmon energy levels, the dispersive shift of the readout mode will vary in time as $\n$ drifts.
At values of $\n$ away from the degeneracy point $\n(\mathrm{mod}~1/2)=\nlb1/4$, the dispersive shifts corresponding to even- and odd-charge-parity will be distinguishable.
With the aid of a quantum-limited Josephson parametric converter~\cite{Bergeal2010}, the rf readout signal was amplified such that the state of the OCS transmon could be detected with high fidelity in a single shot.
For the measurements presented in this paper, we probed the readout resonator at~$\omega_\mathrm{ro}/2\pi = 9.20178~\GHz$ with an integration time per shot of $4.16~\us$.
The average number of photons occupying the readout mode during measurement was~$\approx 10$.
We characterized this readout scheme as a function of time and measured the timescales associated with $\n$ drifts and charge-parity fluctuations.
This simple experiment was composed of three steps: 
\begin{enumerate}
  \item A Ramsey interference experiment was performed to determine the instantaneous~$\n$. Specifically, we  measured~$\delta \omega_{01}(\n)=\nolinebreak\delta \omega_{01}(0)\cos(2\pi\n)$~\cite{Koch2007}, the detuning of $\omega_{01,e}$ from $\overline{\omega_{01}}$.
  \item We acquired $2\times10^4$ high-fidelity dispersive-measurement shots which determine the state of the transmon at a repetition rate of $5~\kHz$.
  \item We repeated step 2, but with each shot preceded by a microwave pulse with carrier frequency $\overline{\omega_{01}}$ to ``scramble" the transmon state, transferring some population from $\ket{0,p}$ to $\ket{1,p}$.
\end{enumerate}
This protocol was repeated 500 times, once every~$40~\s$.
Pulses addressing the transmon had a Gaussian envelope with a carrier frequency $\overline{\omega_{01}}$, which was equally detuned from $\omega_{01,o}$ and $\omega_{01,e}$ at all values of $\n$ so as to be charge-parity insensitive. 
The width of this Gaussian envelope was chosen to be $20~\ns$ to avoid driving the 1-2 transition. 
We refer to these as ``scrambling" pulses because they produced inefficient rotation of the qubit due to the large charge dispersion $\delta\omega_{01}(\n)$.
We note that due to symmetry of the transition spectrum about~$\n=0$ and the degeneracy point~$\n=0.25$, the Ramsey measurement maps all values of $\n$ into the ``half-Brillouin zone"~$[0, 1/4]\subset \mathbb{R}$. 
Thus, we will restrict our discussion of $\n$ to that range.
Below we will describe the outcome of this three-step experiment, emphasizing three separate but related results.
\subsection{Single-Shot Readout of Charge Parity}
\indent Fig.~3(a) shows an example histogram of $2\times10^4$ measurement shots (step 2 of the experiment), where two equally weighted distributions are visible (a histogram of the data projected onto the $I_m$-axis is plotted in~Fig.~4(b)).
The shots in the histogram of Fig.~3(b) were obtained after applying a scrambling pulse to the qubit~(step 3), resulting in four visible distributions. Prior to acquiring these two histograms, a Ramsey measurement~(step 1) was performed to determine that~$n_g=\nlb0.11$.
Each instance of this protocol gave us the readout signal in equilibrium and with scrambled qubit population as a function of $\n$ as it varied in time. 
Fig.~3(c) shows histograms of the phase of the readout signal (step 3) sorted by $\n$ as determined from step 1. The solid lines denote the expected phase for each $\chi_{i,p}$, according to the theory presented earlier and assuming a perfectly reflected signal from an overcoupled resonator~\cite{Pozar2004}:
\begin{equation}
    S_{11}^{i,p}(\omega) = \frac{\omega-\left[\omega_r+\chi_{i,p}(n_\mathrm{g})\right]+i\kappa/2}{\omega-\left[\omega_r+\chi_{i,p}(n_\mathrm{g})\right]-i\kappa/2}.
\end{equation}
Here, $S_{11}^{i,p}(\omega)$ is the frequency-dependent reflection coefficient~[Appendix~B], and the measured phase is given by~$\mathrm{Arg}[S_{11}^{i,p}(\omega_\mathrm{ro})]$.
For our calculation,  we fixed~$\omega_r/2\pi =\nlb9.1979~\GHz$ to match the cavity frequency measured at high probe power ($\approx1~\mathrm{nW}$ at the input of the cavity), beyond the point at which the transmon and readout modes have decoupled~\cite{Reed2010,Verney2019}. 
In our device, the readout mode linewidth~$\kappa/2\pi=\nlb2.5~\MHz$.
The dispersive shifts~$\chi_{i,p}(\n)$ are computed from Eq.~3, where~$g/2\pi=\nlb40~\MHz$ was chosen to match the data.
The charge matrix elements and transition frequencies~$\omega_{ij,p}$ were obtained from numerical simulation~[Fig.~2].
This analysis allows us to confidently assign a joint plasmon and charge-parity state to each distribution in the measurement histogram when~$|\n|\lesssim\nlb0.22$~($\sim\nlb90\%$ of the range). The calculated values of~$\chi_{i,p}(\n)$ can be found in~Fig.~9.\\
\indent Our use of second-order perturbation theory~[Eq.~3] is justified by numerical simulations, which show that the perturbation of the OCS-transmon eigenstates due to the coupling to the readout mode is small over the majority of the~$\n$~range when the number of photons in the readout mode is~$\lesssim10$. 
The wavefunction overlap between the coupled and uncoupled transmon is~$>95\%$, except in the range $0.125\leq \n \leq 0.126$ for the ground state, and when $0.032\leq\n\leq0.034$ or $0.091\leq \n \leq 0.109$ for the excited state.
For example, this approximation breaks down when $\omega_{14,o}$ crosses the bare readout frequency and a more sophisticated theory would need to be employed~\cite{Zhu2013}.
We can thus use simple dispersive readout to probe charge-parity correlations over the majority of $\n$ configurations, and in the next section we will consider the equilibrium case [Fig.~3(a)] where transitions between $\ket{0,e}$ and $\ket{0,o}$ directly measure charge-parity switches.
\subsection{Charge-Parity Dynamics}
In contrast to previous works studying QP dynamics that required coherent operations to map the charge parity of an OCS transmon onto its plasmon eigenstate~\cite{Riste2013,Serniak2018}, here we use our direct readout scheme to track the charge-parity as a function of time.
In~Step~2 of the experiment described above, we measured the OCS transmon state as a function of time with readout parameters that discriminated between the states~$\ket{0,e}$ and~$\ket{0,o}$ (a portion of which is shown in~Fig.~4(a)) and applied a single-threshold (black dashed line) state assignment (red and blue denote~$\geven$ and~$\godd$, respectively) of the charge parity.
This threshold was determined by fitting the distribution of measurement outcomes projected onto the $I_m$-axis to a sum of two Gaussian distributions and taking the midpoint~[Fig.~3(b)].
Here we ignore the residual excited-state population~$\mathcal{P}_1^\mathrm{eq} =\nlb 0.014\pm\nlb0.002$, corresponding to an effective temperature of~$\sim\nlb40~\mK$, which is close to the base temperature of our cryostat~($\approx 20~\mK$).\\
\begin{figure}
	\centering
    \includegraphics[width=\columnwidth]{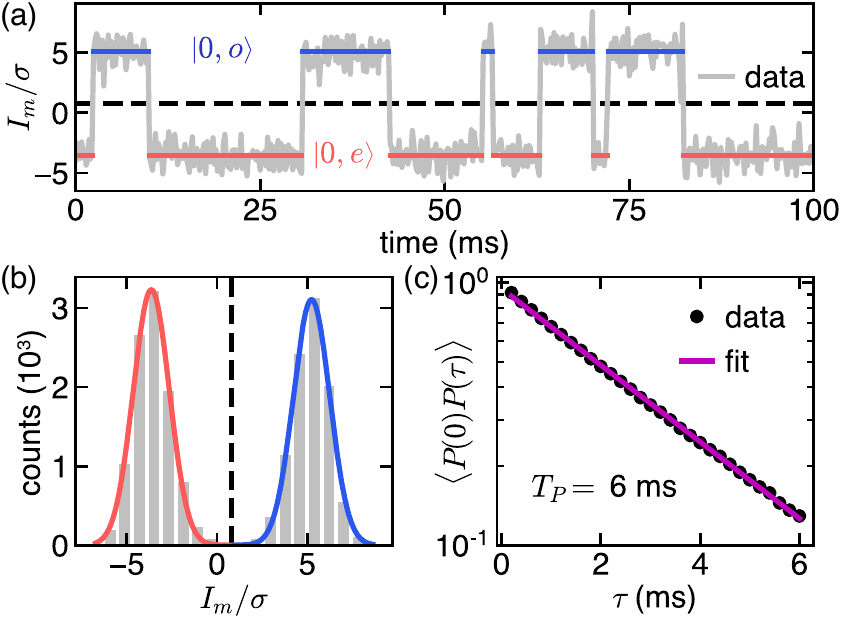}
	\caption{Charge-parity jumps in an OCS transmon. (a) Snapshot of a $\sim4~\mathrm{s}$ time trace from the same data as in Fig.~3(a) projected onto the $I_m$~axis~(grey).
	Charge-parity assignments (red and blue) within the ground-state manifold are obtained with a single threshold at the black-dashed line.
    (b) Histogram of all of the measurements from Fig.~3(a) fit to a sum of two Gaussian distributions, where the colors denote charge-parity assignment. (c) Charge-parity autocorrelation function $\pcorr$ computed from the time trace partially shown in (a) with an exponential fit.
    }
\end{figure}
\indent Having measured the charge parity $P(t)$ of the transmon as a function of time and assuming stationarity and ergodicity, we can compute by a sliding average the charge-parity autocorrelation function
\begin{equation}
    \pcorr = \mathcal{F}^2e^{-2\tau/T_P}.
\end{equation}
For consistency with previous literature, we have defined the charge-parity lifetime $T_P$ as the characteristic time \emph{between charge-parity switches}.
This is a factor of two larger than the timescale for the decay of charge-parity correlations, which is due to equal even-odd and odd-even switching rates.
In this instance where~$\n=0.11$, the fidelity of the charge-parity measurement~$\mathcal{F}\approx0.99$, though this varies with~$\n$ as the two measurement distributions become indistinguishable when~$\n$ approaches the degeneracy point~$\n = 0.25$.
An exponential fit of~$\pcorr$~[Fig.~4(c)] yields~$T_P\approx 6~\ms$, almost an order of magnitude greater than previously reported in~Ref.~\cite{Riste2013} and almost two orders of magnitude greater than in our previous report~Ref.~\cite{Serniak2018}.\\
\indent We attribute this improvement of $T_P$ to additional high-frequency filtering on the input/output line connected to our OCS transmon-cavity system.
The added filter is a~$1~\cm$-long coaxial line filled with Eccosorb CR-110 high-frequency absorber~\cite{Halpern1986}, designed to present an impedance of $50~\Omega$ in the range $2-10~\GHz$~\cite{Pop2014}.
Empirically, placing the filter inside of the sample shielding [Appendix A] is crucial to reducing QP-generating radiation at energies greater than $2\Delta$, the pair-breaking energy.
Further studies to understand this effect and the source of high-frequency, QP-generating radiation are ongoing.
We note that these results are consistent with the notions presented in Ref.~\cite{Houzet2019}, which identified that high-frequency photons could be directly responsible for the observed charge-parity transitions in OCS transmons via photon-assisted QP generation and tunneling processes.\\
\begin{figure}
	\centering
    \includegraphics[width=\columnwidth]{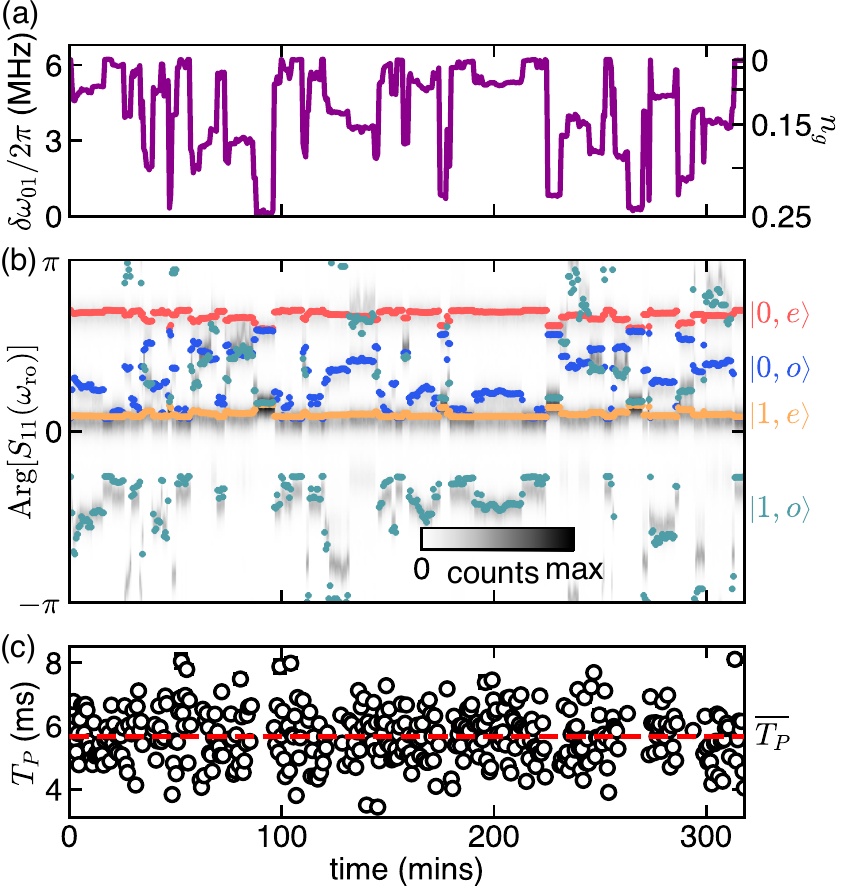}
	\caption{Simultaneous detection of slow and fast charge dynamics.
	(a)~Slow drift of $\n$ probed via a Ramsey experiment (see Ref.~\cite{Serniak2018}).
	The frequency of Ramsey oscillations $\delta \omega_{01}$ is the shift of the qubit transition frequency from its average value $\overline{\omega_{01}}$.
	The right axis converts~$\delta \omega_{01}$ to~$n_g$.
	(b)~Histograms of the phase of repeated dispersive measurements after a state-scrambling pulse [Fig.~3(b)] as a function of time.
	Each instance contains $2\times10^4$ measurement shots acquired immediately after the Ramsey experiment described in (a).
	Colored dots correspond to the predicted phases of each joint plasmon and charge-parity state (labeled on the right) using the theory from the main text, assuming an overcoupled readout resonator.
    (c)~Charge-parity lifetime $T_P$ obtained from the decay of $\pcorr$ as a function of time.
        }
\end{figure}
\subsection{Time Dependence of $T_P$}
\indent The three step experiment was repeated 500 times, the results of which are summarized in Fig.~5.
Ramsey experiments (step 1)~[Fig.~5(a)] determined~$\n$ as a function of time. 
Fig.~5(b) shows histograms of the phase of the readout signal as a function of time, where the overlaid dotted state assignments come from our previous analysis of~$\chi_{i,p}(\n)$ using the measured values of $\n$ in Fig.~5(a).
We compute $\pcorr$ at each of these times [Fig.~5(c)], except in the range~$0.22\lesssim\n\leq\ 0.25$ where the readout distributions corresponding to states $\geven$ and $\godd$ are indistinguishable.
We find an average $\overline{T_P}=5.6~\ms$ with standard deviation $0.8~\ms$.\\
\indent Nonequilibrium QP tunneling will result in a~$T_P$ proportional to~$1/\xqp$. Comparing to the results in Ref.~\cite{Riste2013}~and~\cite{Serniak2018}~(in which both $T_P$ and $\xqp$ are reported), we estimate that the effective residual QP density~$\xqp\sim\nlb10^{-9}$ in this device~\footnote{{Although it is not critical for the analysis presented here, it is worth noting that the definition of $T_P$ in this manuscript, chosen for clarity and convenience, varies slightly from that in previous references~\cite{Riste2013,Serniak2018}. This stems from the fact that the rate of charge-parity switches depends on the plasmon state of the OCS transmon. Here, we report a timescale that is dominated by the rate of transitions between~$\ket{0,o}$ and~$\ket{0,e}$ (equivalent to~$\Gamma_{00}^{eo}$ in the aforementioned references), whereas those references quote a~$1/T_P$ that is effectively the average of this and the rate between~$\ket{1,o}$ and~$\ket{1,e}$.}},
which to the best of our knowledge is the lowest reported value for similar devices.
We find no discernible correlation in~$T_P$ as a function of time, though in this experiment we are only sensitive on the minute timescale.
This sampling rate is limited by the interleaved Ramsey experiment (step 1) and could trivially be increased to $\approx1~\mathrm{Hz}$, at which point more information could be extracted about the spectrum of QP density fluctuations~\cite{Grunhaupt2018}.  
There is also no dependence of~$T_P$ on~$\n$, which is not surprising since $\delta\epsilon_0/k_B\ll20~\mK$, the base temperature of our dilution refrigerator.\\
\begin{figure}
	\centering
    \includegraphics[width=\columnwidth]{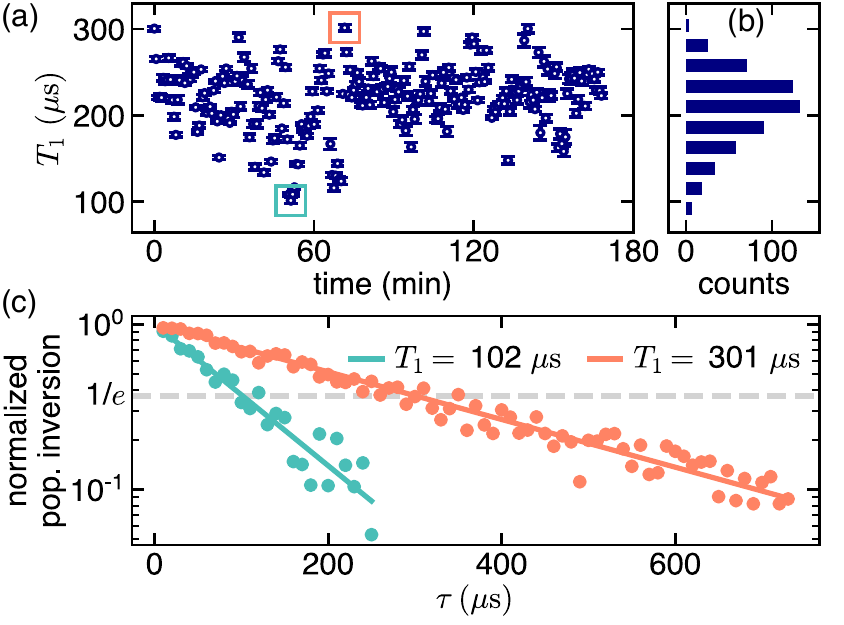} 
	\caption{Fluctuations of OCS transmon energy relaxation time. (a) Relaxation time $T_1$ of the OCS transmon device sampled every $\sim40~\mathrm{s}$.
	(b) Histogram of all $T_1$ measurements (including others not shown in (a)), where the average~$\overline{T_1}=207~\us$.
	(c) Data and fits from the two extremal $T_1$ measurements in (a), marked with green and orange boxes.
    }
\end{figure}
\subsection{Qubit Relaxation and Excitation}
\indent As a further characterization of the sample, we performed standard $T_1$ measurements by applying a scrambling pulse to the qubit and measuring the time it takes for the qubit to thermalize to its equilibrium population distribution in free decay~[Fig.~6]. 
We find that the average $\overline{T_1}\approx207~\us$, but fluctuates in time with a standard deviation of $42~\us$.
At all times, the population decay is well described by a single exponential~[Fig.~6(c)].\\
\indent These results support those in Ref.~\cite{Serniak2018}, which claim that the $T_1$ of this exact device was previously limited to a significant extent by nonequilibrium QPs. 
In that report, we correlated charge-parity transitions with plasmon transitions in an OCS transmon and extracted QP-induced and dielectric-loss-induced transition rates from a fit to a master equation model.
From this, we predicted that if QP-induced dissipation were to be reduced to a negligible level then the transmon would have a residual dielectric quality factor of $\sim4.9\times10^6$ and the equilibrium thermal population of the qubit would be drastically decreased.
Here, with improved rf filtering to reduce QP generation, the measured $\overline{T_1}$ and $\overline{\omega_{01}}$ correspond to a total qubit quality factor of $5.0\times10^6$, extremely close to the predicted ``non-QP" limit.
Suprisingly, we found in Ref.~\cite{Serniak2018} that QP-induced excitation events were the dominant source of residual excited-state population of our OCS transmon. 
We see now that with lower QP density the qubit effective temperature is~$\sim40~\mK$, compared to~$\sim160~\mK$ previously.
These observations indicate that the device was limited in this experiment by dielectric loss~\footnote{This was confirmed qualitatively by qubit-state-conditioned charge-parity autocorrelation measurements similar in theory to those described in Refs.~\cite{Riste2013}~and~\cite{Serniak2018}, but performed with the direct dispersive technique presented in this manuscript. These were performed on a follow-up cooldown and are not presented here.}.
The large fluctuations observed in the measured $T_1$ as a function of time are therefore not due to a fluctuating QP density, but instead to a time dependent coupling to lossy dielectric channels.
Although the source of QP-generating radiation is still unknown, the efficacy of increased filtering at these high frequencies ($\gtrsim100~\GHz$~for our Al-based devices) to reduce QP-induced dissipation is clear.
\section{Discussion and Conclusions}
\indent We have demonstrated a powerful application of OCS-transmon devices through dispersive monitoring of the dynamics of nonequilibrium QPs, which can impair the performance of superconducting quantum circuits.
This technique can be used to extract the rates of all QP-induced qubit transitions as in Refs.~\cite{Riste2013}~and~\cite{Serniak2018}.
We stress that the QP-tunneling rates observed in OCS transmons will be similar to those in traditional high~$E_J/E_C$ transmons by factors of order unity.\\
\indent The observed charge-parity-dependent dispersive shifts of our readout resonator agree well with our simple application of quantum circuit theory~\cite{Vool2017} with the Cooper-pair-box Hamiltonian. 
This strong agreement further supports the idea that the Cooper-pair-box circuit can be used as a testbed for the physics of novel quantum circuit elements. 
Of particular interest are Josephson junctions made from proximity-coupled semiconductors with large spin-orbit coupling and Landé g-factor, which may play host to Majorana fermions when tuned with applied magnetic field into the topological regime~\cite{Lutchyn2010,Oreg2010}.
Proposals suggest embedding these junctions into magnetic-field compatible OCS transmon circuits to look for signatures of this phase transition in spectroscopy experiments~\cite{Ginossar2014,Yavilberg2015}.
These can be observed as changes in transition frequencies or the brightness of certain transitions as a function of~$\n$. In light of our experiments, these features can also be observed in~$\n$-dependent dispersive shifts which are influenced by both the transition frequencies and charge-matrix elements.\\
\indent Additionally, since there is a one-to-one correspondence between the reflected phase indicating~$\ket{0,o}$ and~$\n$, one could use an OCS transmon and the techniques described above as a fast charge sensor with the charge-parity lifetime acting as an upper bound on integration time.
We find the unoptimized charge sensitivity of our OCS-transmon device near $\n=0.11$ to be $\approx\nlb 4.4\times10^{-4}~\mathrm{e}/\sqrt{\mathrm{Hz}}$, which does not change appreciably over the majority of the $\n$ range.
While the rf-SET has better sensitivity to charge fluctuations~\cite{Aassime2001}, the OCS transmon may prove useful for wireless charge sensing with minimal measurement backaction.
Furthermore, our work frames the idea of the ``quantum-capacitance detector"~\cite{Shaw2009,Bueno2010,Stone2012,Echternach2018} in the language of cQED and OCS transmons with symmetric superconducting islands, which may have applications for astronomical detectors.\\
\indent In conclusion, we have achieved direct, dispersive readout of the joint plasmon and charge-parity states of an OCS transmon, i.e.~without performing any coherent operations on the qubit. 
We have demonstrated that, with improved rf filtering, the charge-parity lifetime of typical 3D transmons can be extended to many milliseconds. 
This has also extended the $T_1$ of our OCS transmon to $\approx210~\us$.
Having reduced the effect of nonequilibrium QPs on qubit performance to a negligible level, this provides a clear experimental foundation for further attempts to mitigate other mechanisms of dissipation in superconducting qubits, such as surface dielectric loss~\cite{Wang2015,Dunsworth2017,Calusine2018}.
\section*{Acknowledgements}
We acknowledge insightful discussions with \mbox{Luke} \mbox{Burkhart,} \mbox{Gianluigi Catelani,} \mbox{Gijs de Lange,} \mbox{Leonid} \mbox{Glazman,} \mbox{Manuel} \mbox{Houzet,} \mbox{Dan} \mbox{Prober,} and \mbox{Clarke} \mbox{Smith.}
Facilities use was supported by YINQE and the Yale SEAS cleanroom. 
This research was supported by ARO under Grant No.~W911NF-18-1-0212, and by MURI-ONR under Grant No.~N00014-16-1-2270.
\appendix
\section{Cryogenic Microwave Setup}
\begin{figure}
	\centering
    \includegraphics[width=\columnwidth]{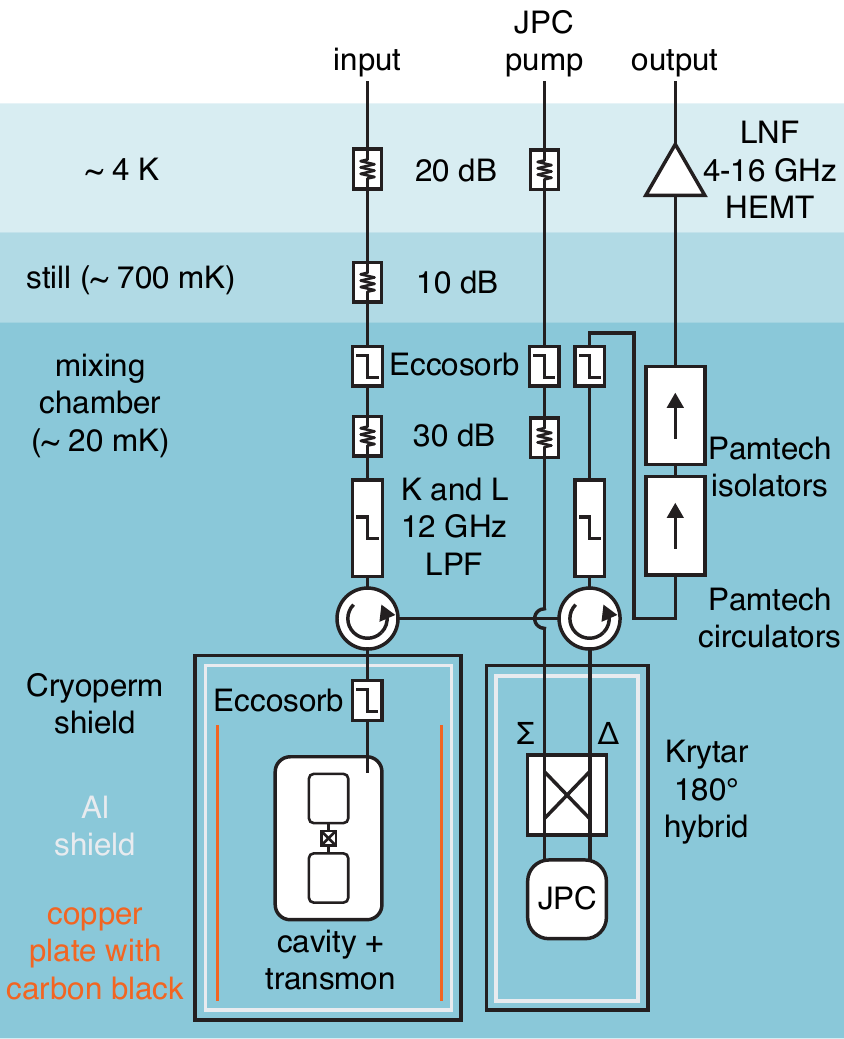} 
	\caption{Wiring diagram of the cryogenic microwave measurement setup.
    }
\end{figure}
The sample was thermalized to the mixing chamber of a cryogen-free dilution refrigerator with base temperature $\approx20~\mK$. The cold rf setup [Fig.~7] was very similar to that of Ref.~\cite{Serniak2018}, with a few modifications, one of which had a direct impact on the improvement of $T_P$. Precisely, this was the addition of an additional Eccosorb CR-110 filter above the input/output port of the OCS transmon-cavity system. We found that placing this filter within the Cryoperm and Al shields was crucial to achieving the largest suppression of QP generation. We note that the coldest radiation shield that is not depicted is thermalized to the still plate (~$\sim 700~\mK$) of the dilution refrigerator.
\section{Frequency-Dependent Phase Response}
\begin{figure}[ht!]
	\centering
    \includegraphics[width=\columnwidth]{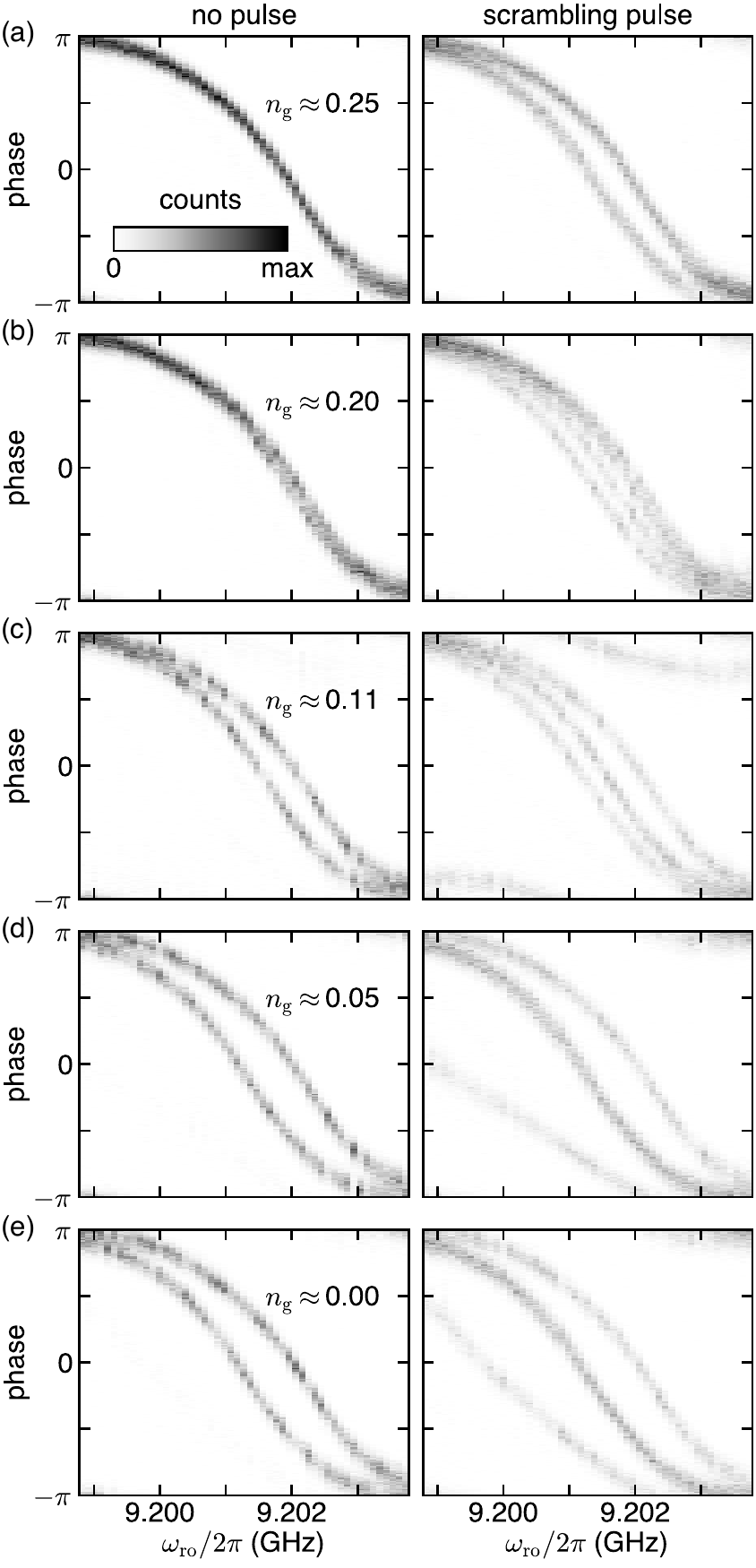} 
	\caption{OCS transmon plasmon- and charge-parity-state dependent readout resonator response. 
	(a-e) Histograms of phase of the signal reflected by the readout resonator as a function of probe frequency at different instances of $\n$.
	The right (left) column is the response with (without) a state-scrambling pulse.
    }
\end{figure}
\begin{figure}
	\centering
    \includegraphics[width=\columnwidth]{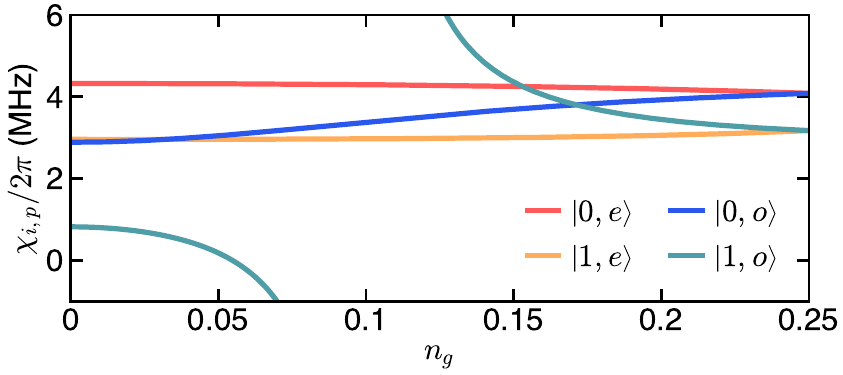} 
	\caption{Theoretically calculated charge-parity-dependent dispersive shifts of the readout mode frequency due to the OCS transmon occupying state $\ket{i,p}$. 
    }
\end{figure}
We performed microwave reflectometry of the single-port readout resonator in the overcoupled regime, in which energy loss through the input/output port is stronger than loss to internal degrees of freedom.
In this regime, the reflection coefficient is characterized by a full $2\pi$ phase roll as a function of frequency with no amplitude response~[Eqn.~4]. 
We resolve this in Fig.~8, where we plot histograms of the measured reflected phase as a function of readout probe frequency, with and without a scrambling pulse preceeding each measurement.
We observe the expected $2\pi$ phase roll for each joint qubit and charge-parity state, which allows for the straightforward extraction of $\chi_{i,p}(\n)$.
The quoted values of $\n$ in each row are obtained by comparison to the data in~Fig.~3(c).
Though there is good agreement with the frequency dependent predictions of Eqn.~4, impedance mismatches within our room-temperature rf-interferometry setup skew these curves.
This contributes a weak background electrical delay to the measured curves.
Operating with a single frequency (as we did for the measurements presented in the main text) avoids this complication.
This technique is particularly convenient for directly observing the charge-parity-dependent dispersive shifts~$\chi_{i,p}(\n)/2\pi$ of the readout mode frequency due to the transmon occupying state~$\ket{i,p}$~[Fig.~9]. Some of the data shown in the main text was acquired at~$\n=0.11$, at which point~$\chi_{1,o}/2\pi\approx11~\MHz$. This is not visible in Fig.~9, in order to better observe the variation of~$\chi_{0,e}$ and~$\chi_{0,o}$ as a function of $\n$.
\section{Device Fabrication}
The OCS transmon was fabricated on a c-plane sapphire wafer. 
The wafer was initially cleaned by sonication in 1-methyl-2-pyrrolidone~(NMP), acetone, and then methanol. 
We then spin coated a bilayer of Microchem 950PMMA A4 on MMA (8.5) MAA EL13 electron-beam-sensitive resists, baking at $\sim\nlb180~^\circ\mathrm{C}$ after each layer. 
After spinning, we sputtered a Au anticharging layer~($\sim\nlb10~\nm$~thick) on the surface.
The transmon pattern was written in a single step with a Raith/Vistec EBPG-5000 100kV electron beam pattern generator. 
After etching away the Au anticharging layer in aqueous KI/I, the pattern was developed in a bath of~3:1~IPA:DI~water at~$6~^\circ\mathrm{C}$.\\
\indent Prior to deposition of Al, an \emph{in~situ} Ar/$\mathrm{O}_2$ ion-beam cleaning was performed in the loadlock of a Plassys UMS-300 evaporation system. 
After a 4~min Ti evaporation (without deposition) to improve the vacuum to~$\approx\nlb5\times\nlb10^{-9}$~Torr, Al junction electrodes (20 and 30 nm thick Al) were deposited at angles of $\pm20^\circ$ in a dedicated evaporation chamber.
Between the Al evaporations, the sample was transferred to a third chamber for thermal oxidation of the first electrode to form the junction barrier. 
This was performed at ambient temperature in a~17:3~Ar:$\mathrm{O}_2$ mixture at a pressure of~30~Torr for~10~min.
To passivate the surface before exposure to air, another thermal oxidation step was performed following the second Al evaporation at~50~Torr for~5~min.
Following the deposition process, the remaining resist and extra Al was removed by a hot NMP liftoff process for one hour with a~30~s sonication step at the end.
A protective layer of Microposit~SC-1827 photoresist was spun and baked at~$\sim80~^\circ\mathrm{C}$ to protect the devices during dicing. 
This protective resist was stripped prior to mounting in the Al 3D readout cavity by sequential rinsing with NMP, acetone, and methanol.
\newpage
\bibliography{ocs_biblio.bib}
\end{document}